# Specification-Driven Development Benchmark: Security Knowledge Transition

Oleg Grynets
EPAM Systems
McLean, Virginia, USA
oleg_grynets@epam.com

Andrii Salyk
EPAM Systems
Lviv, Ukraine
andrii_salyk@epam.com

Vasyl Lyashkevych
EPAM Systems
Lviv, Ukraine
vasyl_lyashkevych@epam.com

Oleh Kaskun
EPAM Systems
Lviv, Ukraine
oleh_kaskun@epam.com

Danyil Zhuravchak
EPAM Systems
Lviv, Ukraine
danyil_zhuravchak@epam.com

***Abstract*—AI-assisted software development is shifting from isolated code completion toward specification-driven generation, where business requirements, technical specifications, and acceptance criteria become operational input for large language model (LLM)-based development agents. This shift creates a security problem: functional behavior is described explicitly, while security behavior remains implicit, generic, or postponed to post-generation review, causing generated systems to satisfy visible functional requirements while failing to preserve authorization rules, ownership boundaries, input validation, token rejection, sensitive data handling, and abuse-case semantics. This paper proposes a security knowledge operationalization approach for AI-assisted specification-driven development, combining two contributions: a Multilayer Specification Security Model that represents security knowledge through traceable relations between system entities, threats, risks, requirements, implementation rules, controls, verification scenarios, and evidence; and a Security Knowledge Transition Method that transforms business and technical specifications into a validated security-enriched generation contract. We evaluate the approach through two empirical studies: one assessing whether an LLM-based pipeline can derive a structured security model under hidden-oracle conditions, and a backend case study in which a GPT-5.3-Codex agent in OpenCode generated FastAPI implementations of a tennis court booking API under three conditions: no explicit security requirements, ASVS-conditioned generation, and Multilayer Security Model conditioning. Evaluated against a hidden 221-test black-box API suite, modal failures decreased from 50 in the baseline to 42 with ASVS and 36 with the Multilayer Security Model, with the strongest improvements in application-specific categories such as business logic and admin safety, providing preliminary evidence that generic security standards improve generated behavior, while application-specific security modeling can further reduce observable security-knowledge loss in spec-driven AI code generation.**

## I. Introduction

LLMs are increasingly used for code generation, refactoring, migration, documentation, test generation, and repository-level assistance. In this setting, the specification is no longer only a human-facing document - it becomes an operational artifact that directly shapes generated code, tests, implementation plans, and downstream verification.

This change is visible in AI-assisted specification-driven development. Instead of a short natural-language request, teams provide structured artifacts: user stories, technical specifications, API contracts, data models, architectural constraints, examples, tests, and runtime instructions. This improves functional controllability because the implementation is constrained by an explicit description of intended behavior.

Security is treated differently. Functional requirements typically define what a system should do: endpoints, schemas, roles, workflows, inputs, outputs, status codes. Security requirements often remain vague or implicit. A specification may state that users can create, update, and cancel bookings, but not whether a user may modify another user's booking. It may define a manager role, but not state whether a manager is restricted to courts they manage. It may require authentication, but omit deactivated-user login rejection, last-admin protection, or token misuse rejection.

This mismatch is methodologically weak. If functional correctness requires explicit specification, security behavior should not be left to model priors. A generation agent may implement basic security patterns because they are common in training data, but this is an unstable outcome of training data, prompt wording, and retrieved context. The result is a form of security knowledge loss: security-relevant knowledge is absent from the specification, too generic to guide implementation, disconnected from verification, or unavailable to the agent at the moment of generation.

This paper treats security knowledge as an explicit lifecycle artifact, consistent with prior work interpreting vulnerability as a lifecycle-relevant characteristic of the software functional state [1], with work on risks in LLM-based development where AI transformations introduce semantic drift and traceability loss [2], and with specification-based Code–Text–Code reengineering, where direct transformation is replaced by an intermediate representation to make generated artifacts controllable, reviewable, and verifiable [3]–[5]. This view is also consistent with evolution-aware specification-driven synthesis, where LLM-based generation is treated as a controlled transformation process requiring monitoring, validation, and adaptation [6], and with multi-agent orchestration approaches where agent involvement and tool selection must be governed explicitly [7].

In this paper, security knowledge transition is defined as the controlled transformation and preservation of security-relevant knowledge across the chain of specification, generation, implementation, and verification artifacts. Formally, let $S$ be a specification bundle, $M$ be a structured security model derived from $S$, $G$ be the generated implementation, and $E$ be the verification evidence obtained through hidden tests and review. Security knowledge transition evaluates whether security obligations represented in $M$ are preserved in $G$ and confirmed by $E$.

Therefore, the problem is not only whether generated code contains vulnerabilities, but whether security intent can be explicitly represented before generation, operationally transferred into code, and verified after generation.

The paper makes four contributions:

- *Multilayer Specification Security Model.* A structured, AI-consumable artifact representing security knowledge through traceable relations among system entities, threats, risks, requirements, implementation rules, controls, verification scenarios, and evidence.
- *Security Knowledge Transition Method*. A method transforming business and technical specifications into a validated security-enriched technical specification for AI-assisted generation.
- *Empirical validation of security-model generation*. Evaluation of whether the pipeline can infer domain-specific security concerns from structured system context under hidden-oracle conditions.
- *Code generation case study*. Evaluation of whether security-requirement conditioning improves AI-generated backends using hidden black-box API tests across three conditions.

Together, these contributions position security knowledge transition as a measurable part of specification-driven AI development rather than as an implicit property of generated code or a purely post-generation review activity.

The central claim is limited but important: generic security standards improve generated security behavior, but they are not sufficient for application-specific security semantics. AI-assisted spec-driven development requires a security artifact layer connecting business rules, system structure, threats, controls, implementation obligations, and verification evidence.

## II. Related Work

### A. LLM-Based Software Development and Code Transformation

Recent LLM-based software engineering research has moved from isolated function-level generation toward repository-level transformation, migration, documentation, and code regeneration. Specification-based Code–Text–Code reengineering proposes that software evolution mediated by LLMs should be treated as a representation-driven transformation process rather than as direct code generation [5]. In this model, source code is transformed into a neutral textual specification that captures program behavior, identifiers, computational flow, side effects, data dependencies, and domain intent. The specification then guides regenerated, migrated, or modified code [5].

This representation-driven view is important for the current paper because security knowledge must also be represented explicitly. If the model receives only functional requirements, it may generate code that passes functional tests while losing authorization assumptions, ownership rules, data sensitivity constraints, or abuse-case rejection behavior. The unified architecture metamodel for information systems developed by generative AI also supports this position by arguing that structured intermediate artifacts improve repeatability and controllability across code, technical documentation, business documentation, and architecture layers [4].

Oracle-to-PostgreSQL migration provides a useful background example of controlled LLM-mediated transformation because it requires preserving syntax, semantics, feature distribution, execution behavior, and data-processing logic. However, in this paper it is used only as a methodological background, while the empirical validation is performed through hidden-oracle security-model generation and backend black-box evaluation. The fine-tuned LLM-based migration framework integrates dataset engineering, feature-aware static analysis, RAG, fine-tuning, error analysis, and evaluation metrics into an iterative workflow [3]. These properties make it suitable for studying whether security-relevant knowledge can also be transferred from specification to generated code.

### B. Secure LLM-Based Code Generation

Pearce et al. demonstrated that GitHub Copilot can generate insecure code in scenarios related to high-risk CWE categories [11]. Perry et al. showed that users assisted by an AI coding assistant wrote less secure code and were more likely to believe that their code was secure [12]. Sandoval et al. and Asare et al. further showed that the security effect of LLM assistance depends on task type, language, user behavior, and interaction pattern [13], [14].

SecurityEval introduced a dataset of 130 samples representing 75 vulnerability types for evaluating machine-learning-based code generation from a security perspective [15]. LLMSecEval and CodeLMSec similarly evaluate generated code under security-relevant prompts and vulnerability-oriented conditions [16], [17].

Security hardening, adversarial testing, and secure-code instruction tuning have also been proposed to improve generated code security [18], [19].

CWEval argues that functionality and security must be evaluated on the same generated artifact rather than in isolation [20]. Dai et al. show that some secure-generation methods appear effective only when security is measured separately, while combined evaluation reveals degraded functionality [21]. BaxBench extends this direction toward backend generation and shows that even correctly generated backend programs may remain exploitable [22].

### C. Specification-Driven Generation and Test-Guided Development

Specification-driven generation is becoming a key direction in AI-assisted software engineering. In embedded software, specification-driven LLM-based generation has been studied using natural-language and formal specifications [23]. Test-driven interactive code generation shows that tests can partially formalize user intent and improve generation outcomes [24]. Recent work also investigates how specification-driven code generation with LLMs can be understood and evaluated empirically [25].

Classical security requirements engineering already emphasizes the importance of misuse cases and structured security requirements [21], [22]. The challenge is to make these security requirements operational in AI-assisted specification-driven workflows.

### D. Secure SDLC Frameworks and Machine-Readable Security Knowledge

Secure SDLC frameworks provide the process-level foundation for this work. OWASP ASVS defines application security requirements and verification criteria [8]. NIST

SSDF frames secure software development practices as activities that should be integrated into the software development lifecycle [9]. OWASP SAMM organizes software assurance activities around lifecycle functions such as Design, Implementation, and Verification [10].

However, these frameworks do not directly solve the AI-native artifact problem. They define what kinds of security work should exist, but they do not define a compact, structured, machine-consumable artifact that can be inserted into an AI-assisted generation contract. ASVS provides reusable security requirements, but real applications require concrete instantiation: which actor, which endpoint, which resource, which trust boundary, which ownership rule, which threat, and which expected result. Therefore, ASVS should be treated as a security knowledge base, not as the final generation artifact.

Threat modeling as code, policy as code, and machine-readable compliance artifacts provide useful building blocks. Code Property Graphs and related program representations also show how structural program knowledge can support vulnerability discovery and program analysis [28]. CodeBERT and GraphCodeBERT demonstrate that code and natural language representations can be connected through learned representations and data-flow-aware code models [29], [30]. Nevertheless, none of these approaches fully defines a specification-to-code security knowledge transition benchmark.

## III. Scientific Gap and Problem Statement

The scientific gap can be stated as follows:

Existing studies show that LLMs can generate insecure code and that prompting, fine-tuning, RAG, and post-generation tests can reduce some weaknesses. However, there is no sufficiently formalized benchmark-oriented method for representing application-specific security knowledge in a specification, transferring this knowledge into generated code, and evaluating whether the transition preserves security intent, implementation obligations, and verification evidence.

This gap contains four sub-problems.

First, there is a security model gap. Secure SDLC frameworks define practices and requirements, but they do not define an AI-consumable specification security model that links system context, threats, risks, requirements, implementation rules, code elements, and tests.

Second, there is a transition gap. Existing research rarely measures how security knowledge moves from business and technical specifications into generated implementation artifacts.

Third, there is a traceability gap. Security requirements are often not connected to threats, assets, endpoints, controls, code fragments, or verification scenarios. Without traceability, it is difficult to determine whether a control is necessary, sufficient, obsolete, or merely decorative.

Fourth, there is a benchmark gap. Many secure-code benchmarks evaluate vulnerabilities after generation, while specification-driven development requires evaluation of the full chain from specification to code and verification. A benchmark should therefore measure not only “is the generated code vulnerable?” but also “was the security knowledge represented, transferred, implemented, and verified?”

The central research question is:

How can security knowledge be represented in specification-driven development, transferred into generated code, and evaluated through a benchmark that separates functional conformance from security behavior?

## IV. Multilayer Specification Security Model

The proposed Multilayer Specification Security Model represents security knowledge as a traceable structure rather than a flat checklist. Fig. 1 summarizes the proposed Multilayer Specification Security Model. The figure should be read from the System Model toward the Verification Layer: system entities define the security context, threats and risks explain why controls are needed, implementation rules specify how obligations enter code, and verification scenarios define how evidence is collected.

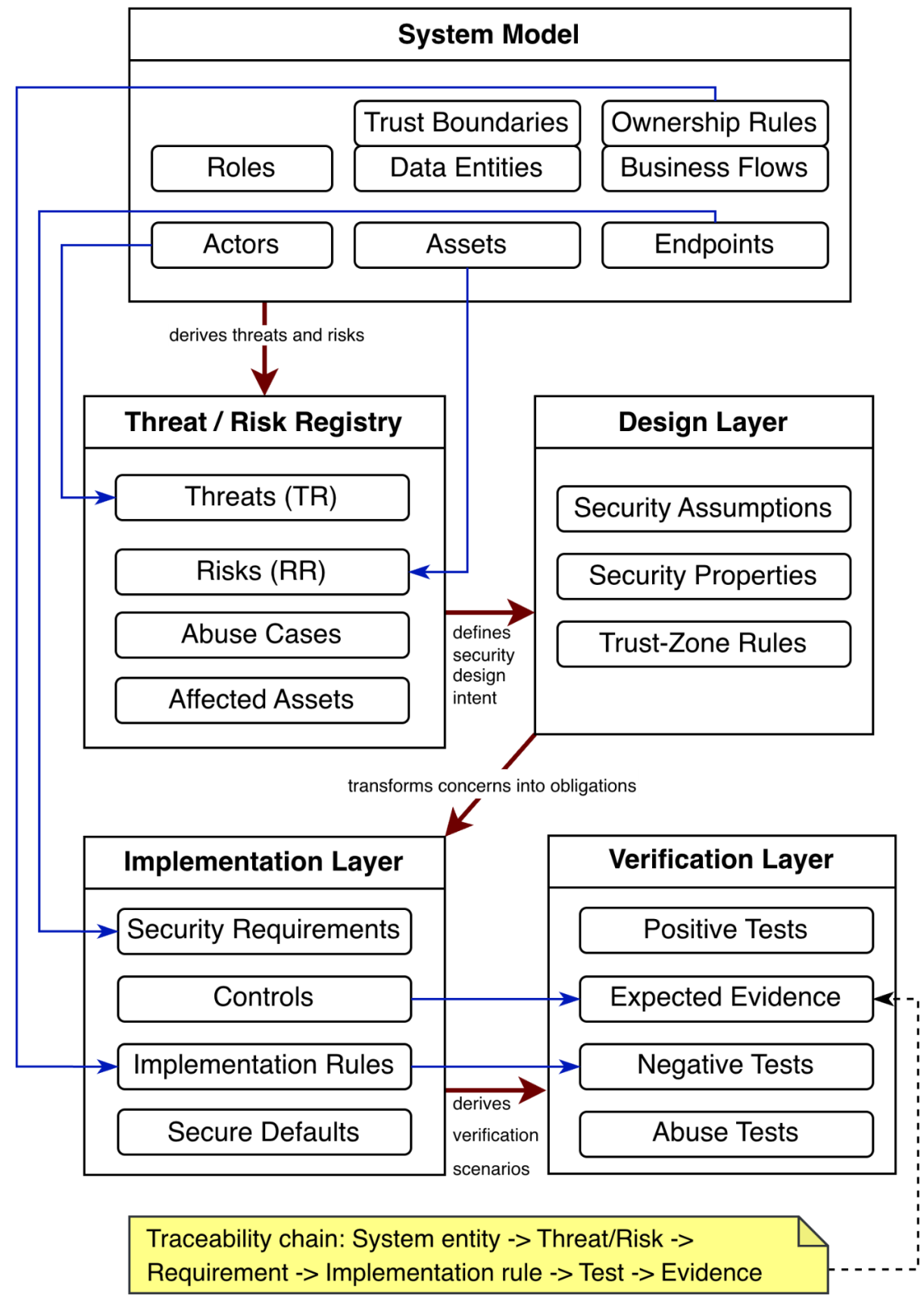

Fig. 1. Multilayer Specification Security Model

Therefore, the model is not only a documentation structure. It is a transition structure: each security element should be traceable from system context to threat, requirement, implementation rule, test, and evidence.

The model contains the following five core elements.

System Model. The System Model contains actors, roles, assets, API endpoints, data entities, components, trust boundaries, dependencies, runtime assumptions, ownership rules, and security-relevant business flows.

Threat/Risk Registry. The Threat/Risk Registry records threats, abuse cases, affected assets, trust boundaries, root causes, expected security properties, mapped control intent, and links to derived requirements and tests.

Design Layer. The Design Layer exposes the security landscape of the system. It answers which assets are protected, which actors interact with the system, which trust boundaries exist, which data flows are sensitive, and which threats are relevant.

Implementation Layer. The Implementation Layer translates threats and design-level security context into actionable security requirements, implementation rules, secure defaults, code constraints, and control obligations.

Verification Layer. The Verification Layer defines how the implementation should be tested. It includes positive tests, negative tests, misuse cases, abuse cases, expected denial behavior, and evidence requirements.

The model is based on the following traceability chain: "*System entity → Threat/Risk → Security requirement → Implementation rule → Code element → Verification scenario → Evidence*".

This structure is essential because security requirements must be defensible. A requirement should not appear merely because it sounds generally secure. It should be traceable to a system fact, a threat or risk, an affected asset, an implementation rule, and a verification obligation.

For example, a generic requirement such as "The system must prevent unauthorized access" is too weak for AI-assisted generation. In the proposed model, it should be instantiated as follows:

A regular user must not read, update, or delete a resource owned by another user unless an administrator role or explicit delegated permission authorizes the operation.

This formulation is more useful because it identifies the actor, resource, operation, ownership constraint, exception condition, and expected forbidden behavior.

The proposed model is not intended to replace OWASP ASVS, OWASP SAMM, or NIST SSDF. Instead, it operationalizes selected secure SDLC concepts into a compact artifact structure suitable for AI-assisted specification-driven development [8]–[10].

## V. Security Knowledge Transition Method

The proposed method transforms business and technical specifications into a security-enriched generation contract. It includes eight stages. Fig. 2 presents the end-to-end Security Knowledge Transition Pipeline. It connects specification validation, security-relevant extraction, model generation, schema validation, expert review, integration into the generation contract, code generation, and hidden evaluation.

The following subsections describe the pipeline stages in detail. The key principle is that security knowledge is generated and validated before code generation, not inferred only after implementation failures are detected.

### A. Specification Acquisition

The method starts with a specification bundle containing business requirements, technical specification, API contract, architecture notes, database schema, runtime contract, and implementation expectations. In API-centric systems, OpenAPI provides a shared contract for endpoint definitions, request and response schemas, authentication schemes, and HTTP-level behavior [31].

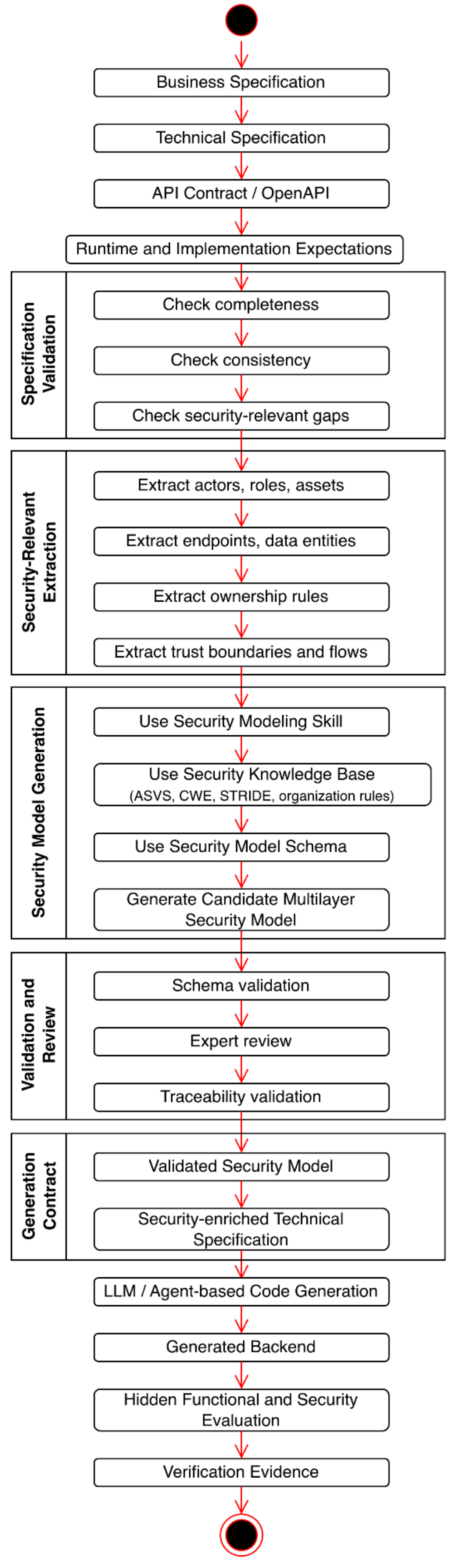

Fig. 2. Security Knowledge Transition Pipeline for Specification-Driven Development

### B. Specification Validation

The business and technical specifications are validated for completeness, consistency, uniqueness, timeliness, and security-relevant adequacy. If actors, ownership rules, roles, permissions, or sensitive resources are missing, the security model may hallucinate assumptions or miss important threats.

This hidden-evaluator configuration separates implementation from assessment and supports a benchmark setting in which generated systems must satisfy the specification rather than memorize evaluator assertions.

### C. Security-Relevant Extraction

The method extracts actors, roles, resources, operations, permissions, ownership rules, sensitive fields, entry points, exit points, trust boundaries, components, dependencies, and runtime assumptions. The output is a normalized System Model.

### D. Threat and Risk Derivation

The method uses the System Model together with security knowledge sources such as OWASP ASVS, CWE, STRIDE, secure coding rules, and organization-specific policies to derive threats and risks [8], [32]. The result is the Threat/Risk Registry.

### E. Security Requirement Synthesis

Threats and risks are transformed into concrete, testable, and traceable security requirements. Each requirement must identify the relevant actor, resource, operation, condition, expected control, and verification obligation.

### F. Implementation Transition

Security requirements are converted into implementation rules. These rules may define endpoint-level authorization, object ownership checks, input validation, database query constraints, secure error behavior, sensitive data handling, token rejection rules, rate-limiting behavior, or secure defaults.

### G. Integration into Final Technical Specification

The validated security model is integrated into the technical specification. The goal is not to append a generic security appendix, but to inject security constraints where the AI development agent will use them: endpoint descriptions, data-model constraints, service behavior, database access rules, implementation expectations, and acceptance criteria.

### H. Verification Transition

The method derives verification scenarios from requirements. These include positive tests, negative tests, misuse tests, abuse tests, and expected evidence. Each test should link back to a requirement and each requirement should link back to a threat or risk.

The output is a security-enriched technical specification that can guide AI-assisted generation and a verification evidence model that can be used to evaluate the generated implementation.

## VI. Benchmark Design

The benchmark evaluates whether a generation agent can produce a backend from a specification bundle while preserving both functional and security expectations. It follows two principles: hidden evaluation and two-axis reporting.

### A. Generator-Visible Artifacts

The generator receives only artifacts that a normal implementer could plausibly receive:

- project overview;
- implementation expectations;
- run contract;
- OpenAPI specification;
- optional security-enriched technical specification;
- optional Multilayer Specification Security Model.

These artifacts define the maximum information available to the generation agent and therefore delimit what can be attributed to specification-driven generation rather than to test leakage.

The generator does not receive hidden tests, scoring rules, evaluator logic, timeout rules, reset procedures, or tolerance annotations.

### B. Hidden Evaluator Artifacts

The evaluator controls:

- functional tests;
- security tests;
- database reset;
- startup checks;
- scoring logic;
- timeout rules;
- failure categorization.

This separation prevents benchmark leakage. Otherwise, generation can collapse into optimization against visible tests rather than implementation from the specification.

### C. Functional Conformance Axis

Functional conformance measures whether the generated backend satisfies the current API contract. It evaluates endpoint availability, request structure, response schemas, status codes, authentication semantics, pagination behavior, and list/detail response differences.

### D. Security Behavior Axis

Security behavior is evaluated using API-level tests informed by OWASP ASVS and restricted to controls observable in a local black-box backend setting [8]. The security suite focuses on authentication enforcement, object-level authorization, property-level authorization, input validation, mass-assignment resistance, token misuse rejection, selected security headers, CORS behavior, cache behavior, and basic abuse resistance.

The benchmark should not collapse functionality and security into one universal score. A backend may satisfy many functional tests while exposing security weaknesses, or satisfy selected security checks while diverging from the API contract. Therefore, the benchmark reports at least two top-level outcomes:

- Functional Conformance Score;
- Security Behavior Score.

This two-axis reporting is important because a backend may be functionally compatible with the API contract while still violating ownership, authorization, validation, or abuse-case constraints.

The security suite is ASVS-informed but does not claim full ASVS coverage. Only controls observable through local black-box API behavior are included. Architecture-level, deployment-level, organizational, and evidence-dependent ASVS requirements are out of scope for this benchmark.

The benchmark package consists of the functional specification bundle, OpenAPI contract, generation instructions, optional ASVS conditioning artifact, optional Multilayer Security Model, generated backend artifacts, hidden test suite, reset scripts, scoring logic, and category-level reports. Only the specification and conditioning artifacts are visible to the generation agent; tests, scoring rules, and failure categories remain hidden until evaluation.

## VII. Experimental Setup

The experimental setup evaluates the proposed Security Knowledge Transition Method from two complementary perspectives. The first study evaluates whether an LLM-based security-model generation pipeline can derive a structured Multilayer Specification Security Model from system context when the original threat and risk model is hidden. The second study evaluates whether providing security knowledge to a code-generation agent improves the runtime behavior of generated backend systems under hidden black-box testing.

The two studies correspond to two different phases of the proposed method. Study 1 evaluates security model production, i.e., whether application-specific security knowledge can be inferred, structured, and represented as a traceable artifact. Study 2 evaluates security model use, i.e., whether this artifact improves generated code when included in the specification-driven generation context.

This design is intentionally limited. Study 1 does not evaluate generated code. It evaluates security-model generation against a hidden oracle. Study 2 does not directly inspect all generated code or formally prove that all security requirements are implemented. It evaluates observable runtime behavior through a hidden black-box API suite. Therefore, the experiments should be interpreted as preliminary evidence of security knowledge transition, not as proof of complete security preservation.

### *A. General Security Knowledge Transition Model*

The general security knowledge transition chain is represented as:

$$S \rightarrow M_s \rightarrow G \rightarrow E. \quad (1)$$

Here, S is the specification input, $M_s$ is the Multilayer Specification Security Model, G is the generated implementation, and E is the verification evidence.

Fig. 3 illustrates the hidden-oracle design of Study 1. The generation agent receives only the visible system context, while the original threats and risks remain hidden and are used only for post-generation evaluation. This setup evaluates inference rather than reproduction. A successful result means that the agent recovered the security concern space from system context, not that it copied the original threat model.

Fig. 4 shows the benchmark workflow used in Study 2. The same functional bundle is provided under three conditioning modes, and each generated backend is evaluated by the same hidden black-box suite. This design isolates the effect of security conditioning. Since the hidden tests and scoring rules are not visible to the generator, differences between conditions can be interpreted as effects of the supplied security knowledge rather than direct test overfitting.

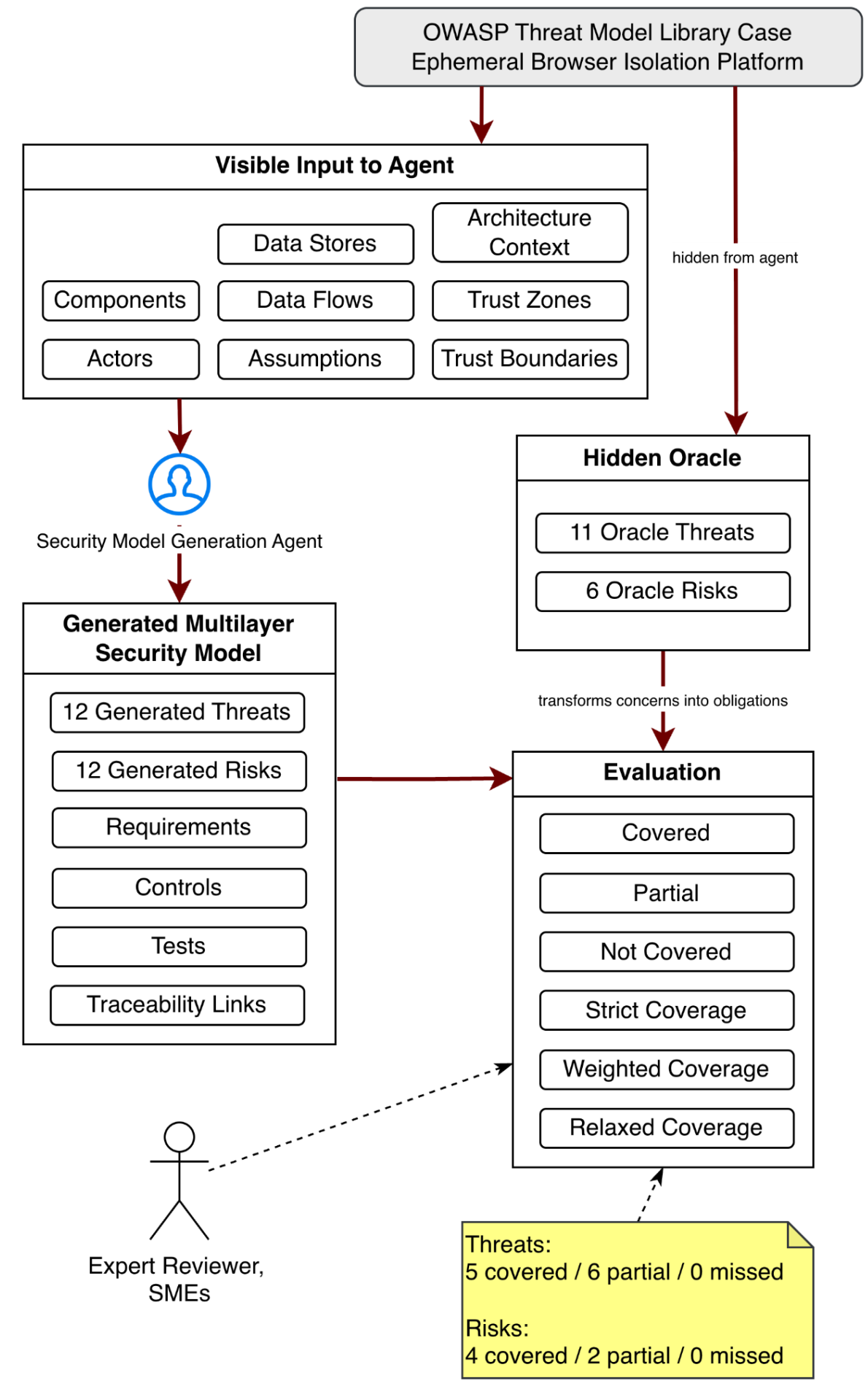

Fig. 3. Hidden-Oracle Security Model Generation Study.

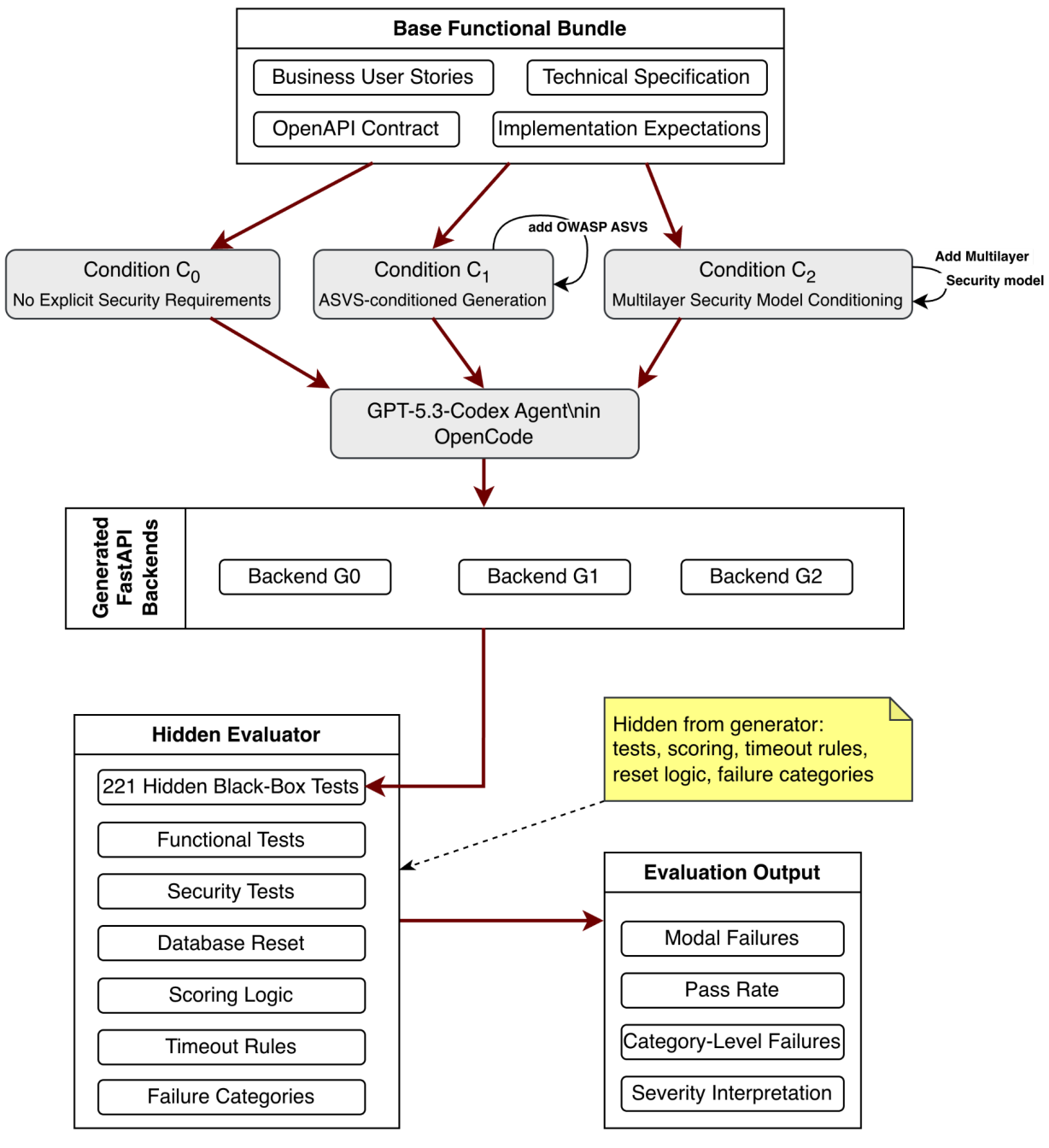

Fig. 4. Backend Generation Benchmark with Hidden Black-Box Evaluation.

However, the two empirical studies instantiate this chain differently. Study 1 evaluates the transition from system context to security model and hidden-oracle comparison. Study 2 evaluates the transition from specification and security conditioning to generated backend behavior under hidden testing. Therefore, the general model should be interpreted as a methodological framework, while the performed experiments evaluate two specific parts of this framework.

### B. Study 1 Formalization: Hidden-Oracle Security Model Generation

In Study 1, the input is not a full implementation specification, it is a structured system-context artifact extracted from the OWASP Threat Model Library case:

$$S_1 = \{A_{ctx}, Z, Bnd, Act, Comp, DS, DF, Assump\}. \quad (2)$$

Here, $A_{ctx}$ is the architecture context, $Z$ is the set of trust zones, $Bnd$ is the set of trust boundaries, $Act$ is the set of actors, $Comp$ is the set of components, $DS$ is the set of data stores, $DF$ is the set of data flows, and $Assump$ is the set of system assumptions.

The generated security model is represented as:

$$M_s^{gen} = \{SM, TR^{gen}, RR^{gen}, Req^{gen}, Ctrl^{gen}, Test^{gen}, L^{gen}\}. \quad (3)$$

Here, $SM$ is the generated system model, $TR^{gen}$ is the generated threat registry, $RR^{gen}$ is the generated risk registry, $Req^{gen}$ is the set of generated security requirements, $Ctrl^{gen}$ is the set of generated controls, $Test^{gen}$ is the set of generated verification scenarios, and $L^{gen}$ is the set of generated traceability links.

The hidden oracle is represented as:

$$O = \{TR_{oracle}, RR_{oracle}\}. \quad (4)$$

Here, $TR_{oracle}$ is the oracle threat set and $RR_{oracle}$ is the oracle risk set.

In the performed experiment, the oracle contained 11 threats and 6 risks:

$$|TR_{oracle}| = 11, \quad |RR_{oracle}| = 6. \quad (5)$$

The generated model contained 12 threats and 12 risks:

$$|TR^{gen}| = 12, \quad |RR^{gen}| = 12. \quad (6)$$

Each oracle item was compared with generated items using semantic matching. The mapping verdict is defined as:

$$map(o_i, g_j) \in \{Covered, Partial, NotCovered\}. \quad (7)$$

Thus, Study 1 evaluates whether the pipeline can recover a useful security concern space from system context. It does not evaluate generated code or runtime behavior.

### C. Study 2 Formalization: Backend Generation with Security Conditioning

In Study 2, the input specification bundle is:

$$S_2 = \{B, T, A, I\}. \quad (8)$$

Here, $B$ is the business-level user story set, $T$ is the technical specification generated through GitHub Spec Kit, $A$ is the OpenAPI contract, and $I$ is the implementation expectation bundle.

The generation condition is represented as:

$$Cond \in \{C_0, C_1, C_2\}. \quad (9)$$

The baseline condition contains only the functional specification bundle:

$$C_0 = S_2. \quad (10)$$

The ASVS-conditioned condition adds OWASP ASVS in machine-readable form:

$$C_1 = S_2 \cup ASVS. \quad (11)$$

The Multilayer Security Model condition adds the application-specific security model:

$$C_2 = S_2 \cup M_s. \quad (12)$$

The generated backend for each condition is represented as:

$$G_k = Agent(C_k), \quad k \in \{0, 1, 2\}. \quad (13)$$

Here, $G_0$ is the backend generated without explicit security requirements, $G_1$ is the backend generated with $ASVS$ conditioning, and $G_2$ is the backend generated with Multilayer Security Model conditioning.

The hidden evaluator is represented as:

$$E = \{T_{hidden}, Reset, Scoring, Timeout, Categories\}. \quad (14)$$

Here, $T_{hidden}$ is the hidden black-box API test suite, $Reset$ is the database reset procedure, $Scoring$ is the scoring logic, $Timeout$ is the timeout policy, and $Categories$ is the failure-category taxonomy.

The generator did not receive hidden tests, scoring logic, timeout rules, or failure categories. This separation prevents direct optimization against evaluator assertions.

The number of hidden black-box tests is: $N = 221$. The modal failure counts obtained in the experiment are: $F_{C_0} = 50, F_{C_1} = 42, F_{C_2} = 36$.

This formalization directly corresponds to the performed experiment: three generation conditions, three generated backend groups, and one hidden black-box evaluation suite.

*D. Hidden Test Suite*

The hidden evaluator controlled functional tests, security tests, database reset, startup checks, scoring, timeouts, and failure categorization. The generator received none of these artifacts. The hidden suite contained 221 black-box items across 15 categories: Admin Safety, Authentication, Authorization, Business Logic, Contract Conformance, CORS, Error Handling, HTTP Methods, Idempotency and Replay, Information Gathering, Input Validation, JWT Token, Schema Strictness, Transport Security, and User Enumeration.

The suite was intentionally API-level and focused on externally observable behavior: authentication enforcement, authorization decisions, ownership constraints, mass-assignment resistance, input validation, token misuse rejection, headers, CORS, cache behavior, and abuse resistance. This scope is consistent with the benchmark goal: evaluating security-relevant runtime behavior that can be observed through local black-box API tests.

*E. Replication Strategy*

Each condition was repeated three times. Because LLM-based agents may produce different implementations across repeated runs, modal reporting was selected to represent the most reproducible outcome for each condition. Failure-count dispersion was calculated using population variance because the three runs constitute the complete replication set for each condition.

The observed failure-count sets were:

$$X_{C_0} = \{50, 50, 48\}. \quad (15)$$

$$X_{C_1} = \{42, 43, 42\}. \quad (16)$$

$$X_{C_2} = \{36, 37, 36\}. \quad (17)$$

The mean failure count is calculated as:

$$\mu = \frac{1}{n}\sum_{i=1}^{n} x_i . \quad (18)$$

The population variance is calculated as:

$$Var(X) = \frac{1}{n}\sum_{i=1}^{n} (x_i - \mu)^2. \quad (19)$$

The standard deviation is calculated as:

$$SD = \sqrt{Var(X)}. \quad (20)$$

The normalized standard deviation is calculated as:

$$SD_{norm} = \frac{NS}{D}. \quad (21)$$

Here, $n = 3,\ N = 221$.

For the baseline condition, the mean failure count is:

$$\mu_{C_0} = \frac{(350+50+48)}{3} = 49.33 . \quad (22)$$

The variance and standard deviation for the baseline condition are:

$$Var(X_{C_0}) = 0.89,\ SD_{C_0} = 0.94 . \quad (23)$$

The normalized standard deviation for the baseline condition is:

$$SD^{C_0}_{norm} = \frac{0.94}{221} = 0.0043 = 0.43\% . \quad (24)$$

The largest observed standard deviation was 0.94 failed tests, or approximately 0.43 % of the 221-test suite. Since the differences between the modal failure counts are 8 tests for $ASVS$ and 14 tests for the Multilayer Security Model, the observed differences substantially exceed run-level stochastic variation.

Modal reporting was used because the goal was to compare the most reproducible generated behavior under each condition. Mean values are still reported to characterize dispersion, but modal failures are used as the primary comparison point because they correspond to the most frequently observed outcome across the three repeated generations.

# VIII. EVALUATION PROTOCOL AND METRICS

The evaluation protocol distinguishes between metrics that are actually computed in the current empirical study and metrics that define the broader benchmark protocol for future extensions.

The current empirical study computes:

- hidden-oracle threat and risk coverage for Study 1;

- failure counts, pass rates, relative failure reductions, run-level dispersion, and category-level reductions for Study 2.

The broader benchmark protocol defines additional metrics such as Security Knowledge Coverage, Requirement Preservation, Implementation Rule Realization, Verification Alignment, and Traceability Coverage. These additional metrics are conceptually valid but are not fully computed in the current experiment unless explicit requirement-to-code and requirement-to-test mappings are available.

### A. Hidden-Oracle Coverage Metrics

For Study 1, the generated security model was compared against the hidden oracle using three coverage measures.

Strict coverage counts only fully covered oracle items:

$$C_{strict} = \frac{N_{covered}}{N_{oracle}} . \tag{25}$$

Weighted coverage gives half credit to partially covered oracle items:

$$C_{weighted} = \frac{N_{covered} + 0.5 \cdot N_{partial}}{N_{oracle}} . \tag{26}$$

Relaxed coverage counts both fully and partially covered items:

$$C_{relaxed} = \frac{N_{covered} + N_{partial}}{N_{oracle}} . \tag{27}$$

Here, $N_{covered}$ is the number of fully covered oracle items, $N_{partial}$ is the number of partially covered oracle items, and $N_{oracle}$ is the total number of oracle items.

For threats, the values are:

$$C^{threat}_{strict} = \frac{5}{11} = 45.45\% . \tag{28}$$

$$C^{threat}_{weighted} = \frac{5+0.5 \cdot 6}{11} = 72.73\% . \tag{29}$$

$$C^{threat}_{relaxed} = \frac{5+6}{11} = 100\% . \tag{30}$$

For risks, the values are:

$$C^{risk}_{strict} = \frac{4}{6} = 66.67\% . \tag{31}$$

$$C^{risk}_{weighted} = \frac{4+0.5 \cdot 2}{6} = 83.33\% . \tag{32}$$

$$C^{risk}_{relaxed} = \frac{4+2}{6} = 100\% . \tag{33}$$

These metrics show that the generated model recovered all oracle threats and risks at least partially, but it did not always preserve the same taxonomy or abstraction level as the hidden oracle.

### B. Backend Runtime Metrics

For Study 2, the main aggregate metrics are failure rate, pass rate, and relative failure reduction.

The failure rate is defined as:

$$FR = \frac{F}{N} . \tag{34}$$

Here, $F$ is the number of failed tests and $N$ is the total number of tests.

The pass rate is defined as:

$$PR = \frac{N-F}{N} . \tag{35}$$

The pass rates for the three conditions are:

$$PR_{C_0} = \frac{221-50}{221} = \frac{171}{221} = 77.38\% . \tag{36}$$

$$PR_{C_1} = \frac{221-42}{221} = \frac{179}{221} = 81.00\% . \tag{37}$$

$$PR_{C_2} = \frac{221-36}{221} = \frac{185}{221} = 83.71\% . \tag{38}$$

The relative failure reduction from condition $A$ to condition $B$ is defined as:

$$RFR(A \rightarrow B) = \frac{F_A - F_B}{F_A} . \tag{39}$$

For ASVS-conditioned generation compared with the baseline:

$$RFR(C_0 \rightarrow C_1) = \frac{50-42}{50} = 16.00\% . \tag{40}$$

For Multilayer Security Model conditioning compared with the baseline:

$$RFR(C_0 \rightarrow C_2) = \frac{50-36}{50} = 28.00\% . \tag{41}$$

For Multilayer Security Model conditioning compared with ASVS:

$$RFR(C_1 \rightarrow C_2) = \frac{42-36}{42} = 14.29\% . \tag{42}$$

These values directly correspond to the obtained experimental results.

### C. Black-Box Approximation of Security Knowledge Preservation

The previous general definition of security knowledge preservation assumed that implemented security knowledge could be directly observed in generated code. That assumption is too strong for the current experiment because Study 2 uses hidden black-box API tests rather than complete white-box code inspection.

Therefore, the observable preserved security knowledge is defined as:

$$K_{obs} = \left\{ k_i \mid k_i \, is \, confirmed \right\} . \quad (43)$$

The black-box approximation of Security Knowledge Preservation is:

$$SKP_{bb} = \frac{|K_{obs}|}{|K_{expected}|} . \quad (44)$$

The black-box approximation of Security Knowledge Loss Rate is:

$$SKLR_{bb} = 1 - SKP_{bb} . \quad (45)$$

In the current empirical study, $SKP_{bb}$ and $SKLR_{bb}$ are not computed as independent metrics because the hidden tests are reported by failure categories rather than by complete security-knowledge element mappings. Instead, hidden failures, category-level failures, and severity classes are used as observable evidence of possible security knowledge transition loss.

Thus, the correct interpretation is:

*Hidden black-box failures are observable evidence of possible security knowledge transition loss under the tested API-level behavior model, but they are not a complete measurement of all security knowledge loss because no full white-box requirement-to-code mapping was performed.*

This distinction prevents overclaiming and keeps the interpretation aligned with the actual experimental design.

### D. Protocol Metrics for Future Benchmark Extensions

The following metrics define the broader benchmark protocol. They are scientifically useful, but in the current paper they should be presented as future or protocol-level metrics unless explicit requirement-to-code and requirement-to-test mappings are provided.

Security Knowledge Coverage measures how much required security knowledge is represented in the generated security model:

$$SKC = \frac{|K_{presented}|}{|K_{required}|} . \quad (46)$$

Security Requirement Preservation measures how much specified security requirement knowledge is preserved in implementation:

$$SRP = \frac{|R_{implemented}|}{|R_{specified}|} . \quad (47)$$

Security Knowledge Loss Rate based on requirement preservation is:

$$SKLR_R = 1 - SRP . \quad (48)$$

Implementation Rule Realization measures how many implementation rules are reflected in generated code:

$$IRR = \frac{|IR_{realized}|}{|IR_{specified}|} . \quad (49)$$

Verification Alignment Score measures how well verification scenarios correspond to intended requirements:

$$VAS = \frac{|V_{aligned}|}{|V_{total}|} . \quad (50)$$

Traceability Coverage measures how complete the traceability chain is:

$$TC = \frac{|Links_{valid}|}{|Links_{expected}|} . \quad (51)$$

The current empirical study computes hidden-oracle coverage, failure rates, pass rates, relative failure reductions, run-level dispersion, and category-level reductions. Metrics such as $SKC$, $SRP$, $IRR$, $VAS$, and $TC$ define the broader evaluation protocol and can be fully operationalized in future studies with explicit requirement-to-code and requirement-to-test mappings.

## IX. RESULTS

### A. Study 1: Hidden-Oracle Security Model Generation Results

The hidden oracle contained 11 threats and 6 risks. The generated model contained 12 threats and 12 risks. All oracle threats and risks were at least partially represented in the generated model. Strong coverage was achieved for 5 threats and 4 risks; the remaining oracle items were partially covered. There were no hard misses.

Table I summarizes the hidden-oracle comparison between the original threat/risk model and the generated Multilayer Security Model.

The absence of hard misses indicates that the generated model recovered all oracle concerns at least partially, but the distinction between Covered and Partial shows that exact taxonomy preservation was not always achieved.

TABLE I. HIDDEN-ORACLE SECURITY MODEL GENERATION RESULTS

| Dimension | Result |
|---|---|
| Oracle threats | 11 |
| Generated threats | 12 |
| Threats covered | 5 / 11 |
| Threats partially covered | 6 / 11 |
| Threats not covered | 0 / 11 |
| Oracle risks | 6 |
| Generated risks | 12 |
| Risks covered | 4 / 6 |
| Risks partially covered | 2 / 6 |
| Risks not covered | 0 / 6 |

Using the coverage metrics defined above, the generated model achieved the following values. Table II converts the raw hidden-oracle counts into strict, weighted, and relaxed coverage metrics.

TABLE II. HIDDEN-ORACLE COVERAGE METRICS

| Artifact type | Strict coverage | Weighted coverage | Relaxed coverage |
|---|---|---|---|
| Threats | 5/11 = 45.45% | (5 + 0.5 × 6)/11 = 72.73% | 11/11 = 100.00% |
| Risks | 4/6 = 66.67% | (4 + 0.5 × 2)/6 = 83.33% | 6/6 = 100.00% |

These values show that relaxed coverage is complete, while strict coverage remains moderate. Therefore, the generated model is useful as a first-pass security reasoning artifact but still requires expert review for taxonomy alignment and domain-specific refinement.

The result shows that the generated model recovered the full security concern space at a relaxed level, but did not always reproduce the exact oracle taxonomy. The main weakness was framing rather than recall. For example, control-plane compromise was decomposed into separate platform-security risks such as privilege escalation, audit-log manipulation, TTL bypass, and authentication-service compromise. This decomposition may be useful for downstream implementation, but it weakens direct alignment with the hidden oracle taxonomy.

The most important positive result was the traceability structure. A reviewer could select a requirement and inspect which threat motivated it, which risk it mitigated, which asset it protected, which control implemented it, and which test verified it. This makes the generated model auditable and suitable for downstream use as a security-enriched generation artifact.

## B. Study 2: Run-Level Failure Dispersion

Each experimental condition was repeated three times. Table III reports run-level dispersion across the three repeated generations for each condition.

TABLE III. RUN-LEVEL FAILURE DISPERSION

| Condition | Run failures | Mean | Mode | Var(X) | SD | SD / 221 tests |
|---|---|---|---|---|---|---|
| No explicit security requirements | 50, 50, 48 | 49.33 | 50 | 0.89 | 0.94 | 0.43 % |
| ASVS-conditioned generation | 42, 43, 42 | 42.33 | 42 | 0.22 | 0.47 | 0.21 % |
| Multilayer Security Model | 36, 37, 36 | 36.33 | 36 | 0.22 | 0.47 | 0.21 % |

Since the largest standard deviation is less than one failed test, while the differences between conditions are 8 and 14 failed tests, the observed improvements exceed the measured run-level variability.

The differences between conditions are larger than run-level dispersion. The baseline modal result was 50 failures, ASVS-conditioned generation produced 42 failures, and Multilayer Security Model conditioning produced 36 failures. Therefore, the observed reduction is not explained by stochastic variation between repeated generations.

## C. Aggregate Backend Evaluation Results

The modal failure counts correspond to pass rates of 171/221 = 77.38% for the baseline, 179/221 = 81.00% for ASVS-conditioned generation, and 185/221 = 83.71% for Multilayer Security Model conditioning. Compared with the baseline, ASVS reduced modal failures by 8 tests, corresponding to a 16.00% relative failure reduction. The Multilayer Security Model reduced modal failures by 14 tests, corresponding to a 28.00% relative failure reduction. Compared with ASVS alone, the Multilayer Security Model reduced failures by 6 additional tests, corresponding to a 14.29% relative reduction over ASVS.

Table IV aggregates the modal backend results and reports pass rates and relative failure reductions.

TABLE IV. AGGREGATE MODAL PERFORMANCE ACROSS GENERATION CONDITIONS

| Condition | Modal failures | Passes | Pass rate | Relative failure reduction vs. baseline |
|---|---|---|---|---|
| No explicit security requirements | 50 | 171 | 77.38% | — |
| ASVS-conditioned generation | 42 | 179 | 81.00% | 16.00% |
| Multilayer Security Model | 36 | 185 | 83.71% | 28.00% |

Table IV aggregates the modal backend results and reports pass rates and relative failure reductions. The aggregate results show a monotonic improvement from baseline to ASVS and from ASVS to the Multilayer Security Model. This supports the hypothesis that application-specific security knowledge provides additional value beyond generic security standards. Fig. 5 visualizes the same trend across the three conditioning modes.

Table IV gives the numerical summary of the aggregate backend evaluation, while Fig. 5 visualizes the same trend across the three conditioning modes.

The monotonic decrease in modal failures suggests that additional security knowledge improves generated behavior. However, the improvement is not uniform across all failure categories, which motivates the category-level analysis in the next subsection.

## D. Category-Level Results

The category-level results show that improvements were not uniformly distributed across the suite. The strongest effects appeared in application-specific categories requiring explicit domain-level security reasoning. Table V reports the distribution of modal failures across the 15 hidden test categories.

The distribution shows that most residual failures are concentrated in Business Logic, Contract Conformance, and Input Validation. This indicates that aggregate scores alone are insufficient and should be interpreted together with category-level evidence.

TABLE V. Test-suite size by category and modal category results

| Category | Tests | No security fails | ASVS fails | Security model fails |
|---|---|---|---|---|
| Admin Safety | 4 | 2 | 2 | 0 |
| Authentication | 28 | 1 | 1 | 1 |
| Authorization | 30 | 2 | 0 | 1 |
| Business Logic | 43 | 17 | 15 | 9 |
| Contract Conformance | 34 | 13 | 13 | 14 |
| CORS | 3 | 0 | 0 | 0 |
| Error Handling | 5 | 1 | 0 | 1 |
| HTTP Methods | 9 | 0 | 0 | 0 |
| Idempotency and Replay | 2 | 0 | 0 | 0 |
| Information Gathering | 4 | 0 | 0 | 0 |
| Input Validation | 38 | 9 | 7 | 7 |
| JWT Token | 6 | 1 | 0 | 0 |
| Schema Strictness | 11 | 2 | 2 | 1 |
| Transport Security | 3 | 2 | 2 | 2 |
| User Enumeration | 1 | 0 | 0 | 0 |
| **Total** | **221** | **50** | **42** | **36** |

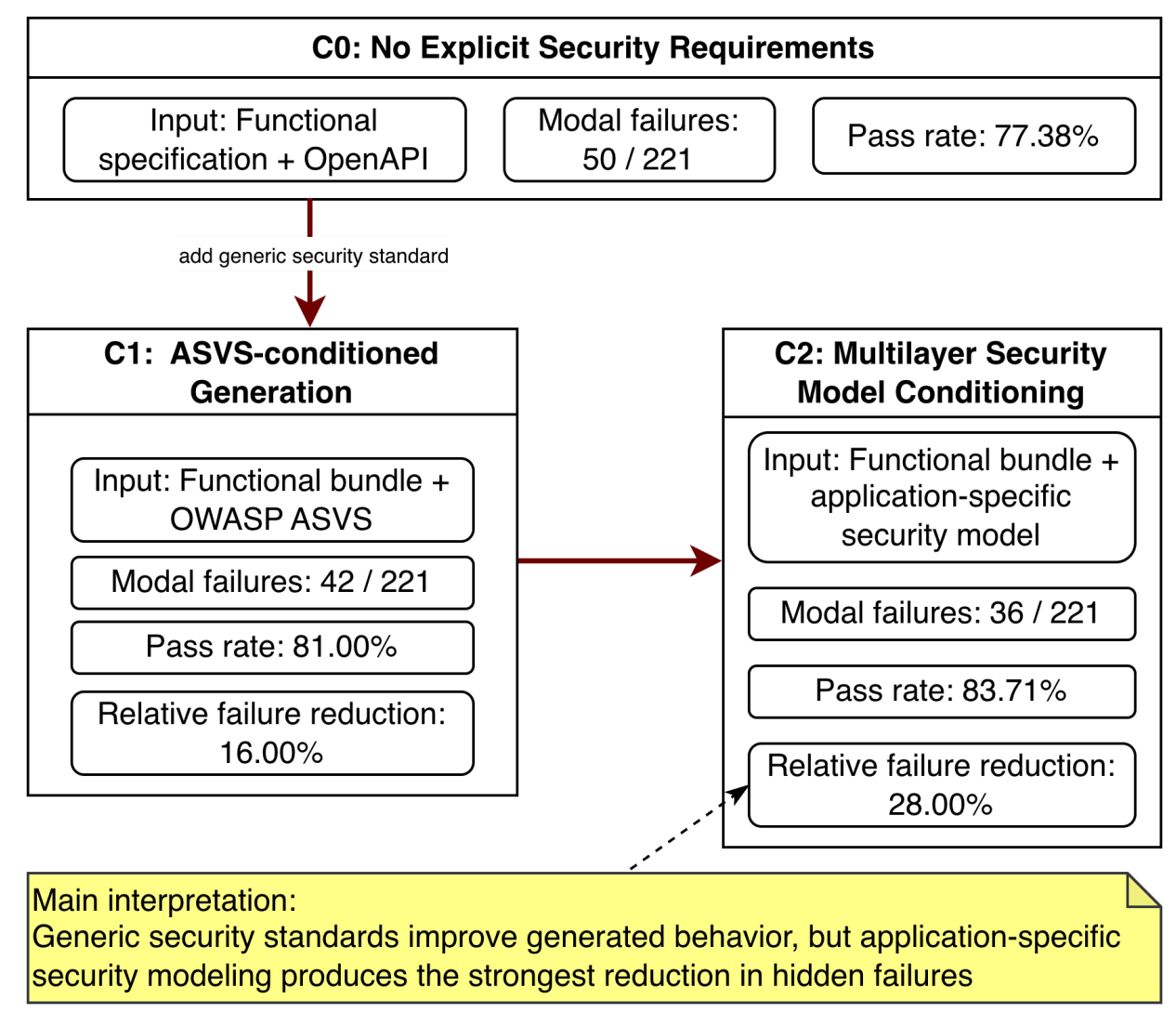

Fig. 5. Experimental Conditions and Aggregate Results.

The strongest improvement appeared in Business Logic, where failures decreased from 17 in the baseline to 9 with the Multilayer Security Model, corresponding to a 47.06% relative reduction. Admin Safety improved from 2 failures to 0, eliminating this category in the modal run. JWT Token failures decreased from 1 to 0, and Schema Strictness improved from 2 to 1. Input Validation improved from 9 to 7 under ASVS, but did not improve further under the security model, suggesting that generic standards already cover part of this class. Transport Security remained unchanged, which is expected because local black-box API evaluation cannot fully enforce deployment-layer controls.

Table VI focuses on the categories with the most relevant changes between the baseline and Multilayer Security Model conditions.

TABLE VI. Category-level improvement from baseline to Multilayer Security Model

| Category | Baseline failures | Security model failures | Absolute reduction | Relative reduction |
|---|---|---|---|---|
| Admin Safety | 2 | 0 | 2 | 100.00% |
| Business Logic | 17 | 9 | 8 | 47.06% |
| Input Validation | 9 | 7 | 2 | 22.22% |
| JWT Token | 1 | 0 | 1 | 100.00% |
| Schema Strictness | 2 | 1 | 1 | 50.00% |
| Contract Conformance | 13 | 14 | -1 | -7.69% |
| Transport Security | 2 | 2 | 0 | 0.00% |

The strongest reductions occur in Admin Safety and Business Logic, which are application-specific categories. This supports the claim that the proposed model is most useful when security requirements depend on domain-specific roles, ownership rules, and state-transition constraints.

The category-level distribution is more informative than the aggregate failure count alone. The main value of the Multilayer Security Model is not simply that it reduces failures, but that it reduces failures in categories where missing security knowledge is application-specific and security-critical.

Table VI isolates the categories in which the Multilayer Security Model changed the result most strongly. Fig. 6 visualizes this category-level effect and separates application-specific improvements from unchanged or partially improved categories.

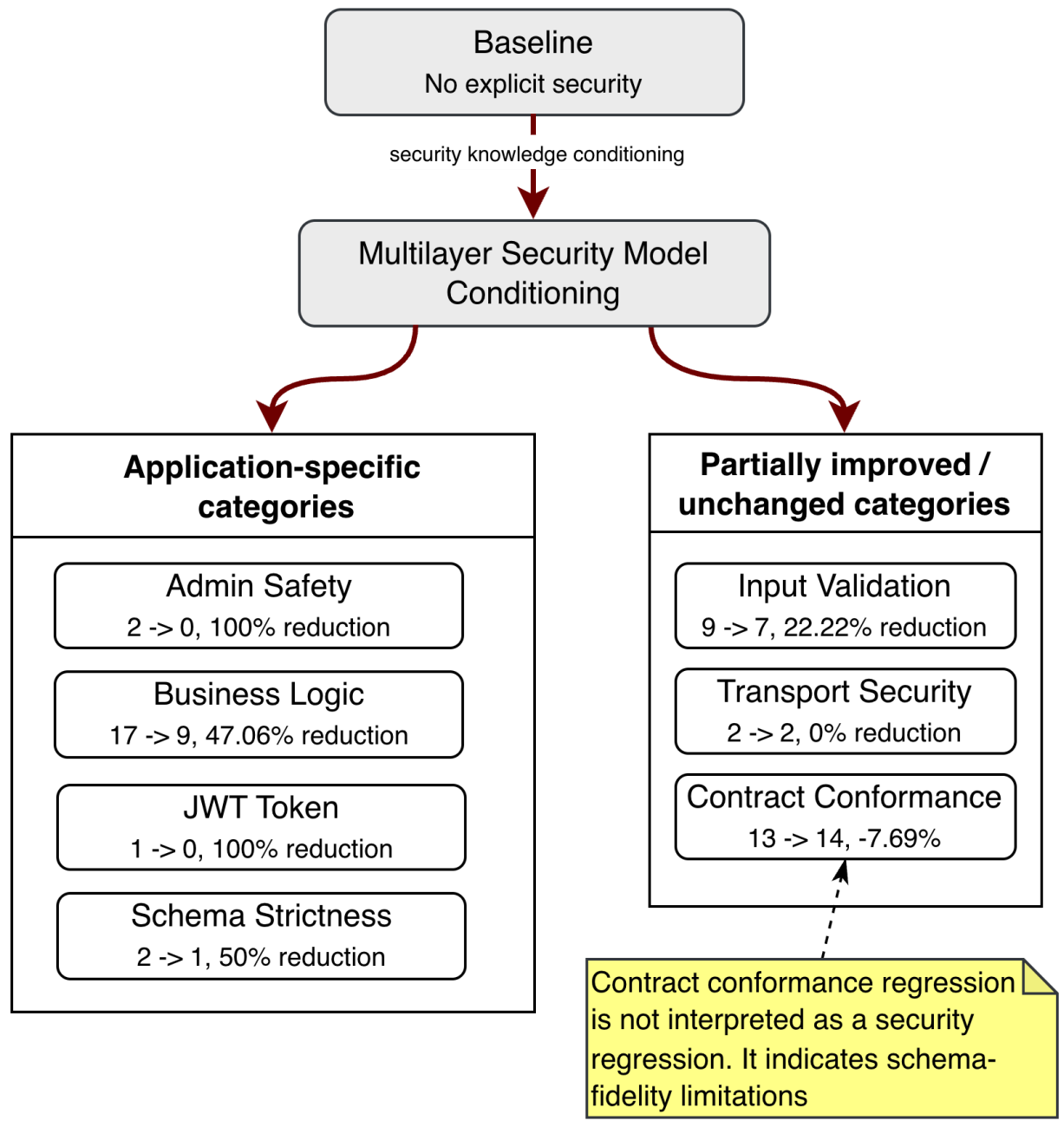

Fig. 6. Category-Level Failure Reduction from Baseline to Security Model.

The figure confirms that the strongest improvements occur where generic standards are weakest: application-specific business rules, admin recoverability, token handling, and schema-sensitive constraints.

## *E. Contract Conformance Regression*

Contract Conformance slightly worsened with the security model condition, increasing from 13 failures to 14 failures.

This should not be interpreted as a security regression. OpenAPI defines the API interface, while the security model optimizes security semantics. These objectives can diverge: a backend may enforce a correct security decision but return a response envelope or status code that does not match the schema. For example, rejecting a dangerous action with a generic error body instead of the expected schema is a contract failure even if the security intent is satisfied.

This result indicates a schema-fidelity limitation of the current pipeline. The natural next step is to combine security-model conditioning with contract-driven validation and post-generation repair. In other words, the security model should not replace OpenAPI conformance; it should be integrated with contract-aware generation and repair.

### F. Severity Interpretation

Table VII classifies the main failure types by severity to avoid treating all failures as equally important.

TABLE VII. SEVERITY CLASSIFICATION OF REAL SECURITY FINDINGS

| Finding type | Severity | Short rationale |
|---|---|---|
| Last admin demotion / admin self-deactivation | High | Admin recoverability failure |
| Deactivated user can log in again | High | Account-lifecycle bypass |
| Missing object/function authorization | High | Access-control bypass |
| Mass assignment of role, owner, or active status | High | Privilege/state manipulation |
| Invalid terminal booking-state transition | Medium | Workflow/audit integrity loss |
| Price snapshot integrity failure | Medium | Financial integrity risk |
| Empty or weak password accepted | Medium | Weak authentication control |
| Invalid dates, UUIDs, names, prices, enums | Low/Medium | Data-integrity or abuse risk |
| Error envelope inconsistency / minor disclosure | Low/Medium | Hardening issue unless internals leak |
| Missing transport-security controls locally | Low/Medium | Deployment-layer dependent |
| Pure OpenAPI schema mismatch | Low | Reliability issue, not necessarily exploitable |

This classification shows that some failures represent reliability or schema-fidelity issues, while others correspond to high-impact security failures. Therefore, severity-aware interpretation is necessary for judging the practical effect of security conditioning.

## X. DISCUSSION

The results provide two complementary forms of evidence. Study 1 evaluates whether the proposed schema and generation pipeline can construct a structured security model from system context. Study 2 evaluates whether such security knowledge improves generated runtime behavior.

Fig. 7 explains how the experimental results should be interpreted. Hidden failures are not treated as a complete proof of security knowledge loss, but as observable runtime evidence that some expected security behavior was not preserved.

This interpretation keeps the claim conservative: the benchmark provides black-box evidence of possible security knowledge transition loss, while full confirmation would require white-box tracing from requirements to implementation and tests.

In Study 1, the generated model achieved complete relaxed coverage: all 11 oracle threats and all 6 oracle risks were at least partially represented. However, strict coverage was lower: 5/11 threats and 4/6 risks were fully covered. Weighted coverage reached 72.73% for threats and 83.33% for risks. This indicates that the pipeline can recover the security concern space, but expert review is required to refine taxonomy, abstraction level, and domain-specific framing.

In Study 2, security conditioning reduced hidden black-box failures in a controlled code generation setting. The baseline condition produced 50 modal failures out of 221 tests, corresponding to a 77.38% pass rate. ASVS-conditioned generation reduced failures to 42, improving the pass rate to 81.00% and reducing failures by 16.00% relative to the baseline. Multilayer Security Model conditioning reduced failures to 36, improving the pass rate to 83.71% and reducing failures by 28.00% relative to the baseline.

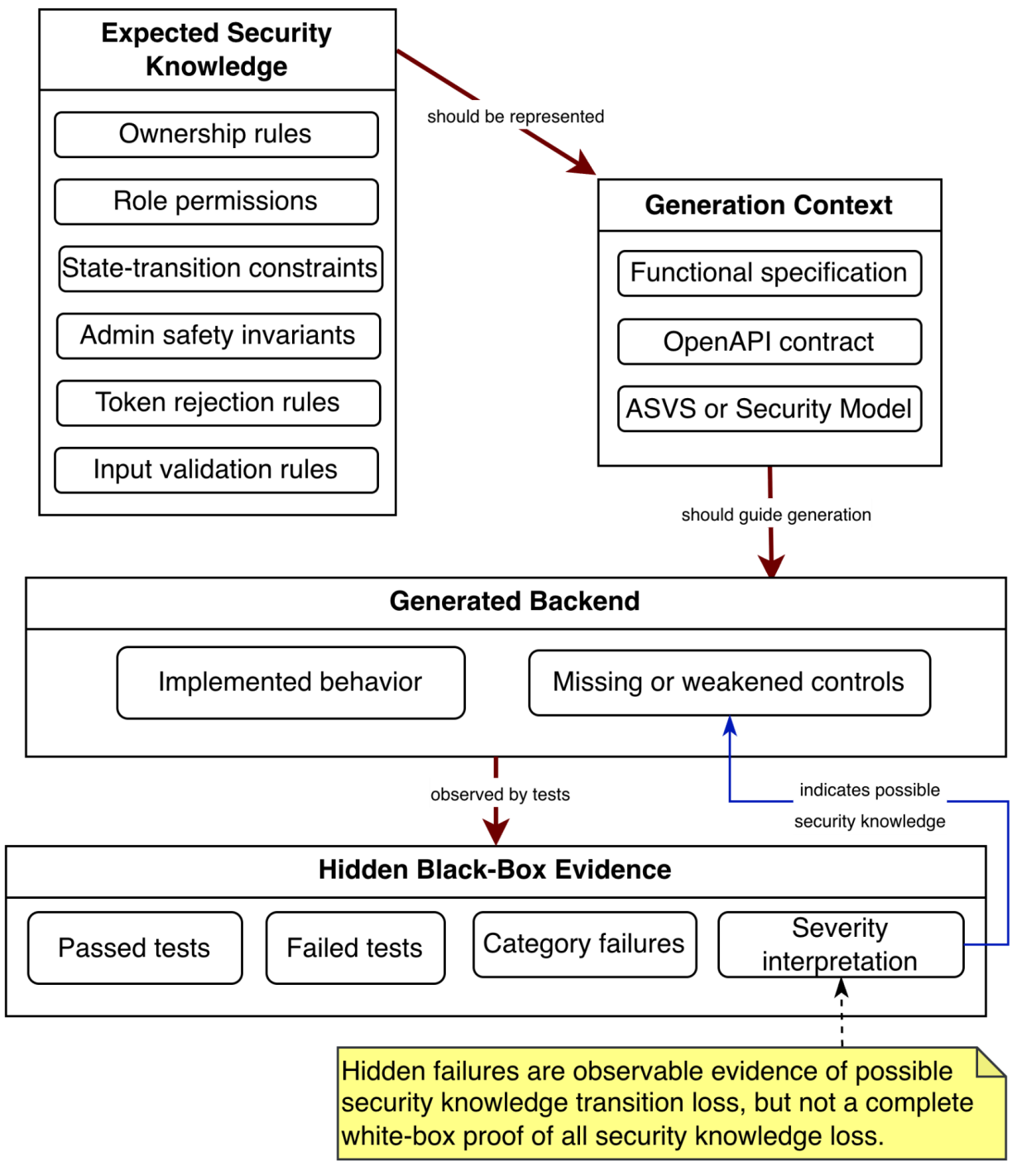

Fig. 7. Security Knowledge Loss Interpretation Model.

The comparison between ASVS-conditioned generation and Multilayer Security Model conditioning clarifies the role of general and application-specific security knowledge. ASVS improves the result because it provides structured generic security expectations: authentication, access control, validation, token handling, error behavior, and secure defaults. However, ASVS does not know the application domain. It does not know that a manager should be restricted to courts they manage, that a last administrator should not be demoted, that a booking must preserve price snapshot integrity, or that terminal booking states should not be reopened. These constraints are not generic security controls; they are domain-specific security invariants. The Multilayer Security Model improves over ASVS because it binds general security concepts to concrete actors, assets, resources, operations, state transitions, and expected denial behaviors.

The results also show that the security model is not a substitute for verification. The best condition still had 36 modal failures out of 221 tests. Therefore, the proposed model should be interpreted as a security-requirement operationalization layer, not as a guarantee of secure code. It improves the generation input, but hidden black-box testing remains necessary for output validation.

Table VIII connects the claimed contributions of the paper with the empirical evidence reported in the two studies.

This mapping clarifies the scope of the evidence. The results support the usefulness of the model and method, but they do not claim complete security assurance or generalization across all domains.

TABLE VIII. RESEARCH CONTRIBUTIONS AND SUPPORTING EVIDENCE

| Contribution | Evidence in the paper | Main result |
|---|---|---|
| Multilayer Specification Security Model | Formal model and hidden-oracle study | All 11 oracle threats and all 6 oracle risks were at least partially recovered |
| Security Knowledge Transition Method | Eight-stage method, schema validation, expert review | Generated security model produced traceable chains from threats to requirements, controls, and tests |
| Benchmark-style evaluation | Hidden 221-test black-box API suite | Tests remained hidden from the generator; evaluation used independent runtime evidence |
| Effect of generic security knowledge | ASVS-conditioned generation | Modal failures decreased from 50 to 42 |
| Effect of application-specific security model | Multilayer Security Model conditioning | Modal failures decreased from 50 to 36; Business Logic failures decreased from 17 to 9 |
| Need for verification | Best condition still failed 36/221 tests | Security model improves generation input but does not replace hidden testing |

The practical implication is that security requirements should be generated and validated before code generation, not treated only as review feedback after implementation. In AI-assisted specification-driven development, the security model acts as an intermediate artifact that translates business rules and threat assumptions into implementation and verification obligations.

Overall, the results support the central claim: generic security standards help, but application-specific security modeling is needed to reduce security knowledge loss in AI-assisted specification-driven development.

## XI. THREATS TO VALIDITY

### A. *Internal Validity*

Three replications per condition are sufficient to observe large differences but not enough to fully characterize stochastic generation behavior. Future work should increase the number of replications, evaluate multiple model families, and explicitly control temperature, agent settings, context window configuration, and tool policies.

### B. *Construct Validity*

Failure counts approximate security quality but do not fully represent exploitability. Some failures are pure OpenAPI contract mismatches, while others are high-impact security failures. The study partially addresses this limitation through category-level results and severity classification, but future work should separate functional failures, schema failures, security failures, and exploitability more explicitly.

### C. *External Validity*

The backend study uses one domain: tennis court booking. The hidden-oracle study uses one browser isolation system. Therefore, the results may not generalize to financial systems, healthcare systems, distributed microservices, embedded systems, or large legacy systems. More domains are required before broad generalization.

### D. *Evaluation Validity*

The security-model generation study used manual semantic matching, which introduces reviewer subjectivity. Future work should use multiple reviewers, explicit mapping guidelines, and inter-rater agreement. In addition, verification scenarios in Study 1 were assessed as traceability artifacts and not executed against a real implementation.

### E. *Benchmark Leakage Validity*

The hidden-test design reduces the risk of direct overfitting because the generation agent did not receive the tests, scoring logic, timeout rules, or failure categories. However, it does not eliminate all forms of indirect leakage. The model may have learned common patterns from public API examples, FastAPI applications, or security testing conventions during pretraining. Therefore, the benchmark should be interpreted as a hidden-evaluator study rather than as a fully leakage-free security proof.

### F. *Oracle Taxonomy Validity*

In Study 1, the generated model was compared against the hidden oracle using semantic matching. Some generated threats were valid but framed differently from the oracle. This creates a taxonomy-alignment problem: strict matching may underestimate useful security reasoning, while relaxed matching may overestimate semantic equivalence. Weighted coverage partially mitigates this issue, but future studies should use multiple expert reviewers and explicit mapping guidelines.

### G. *Tooling Validity*

The backend study used GPT-5.3-Codex in OpenCode generating FastAPI backends. Different LLMs, agent frameworks, programming languages, databases, or prompting policies may produce different behavior.

## CONCLUSION

This paper proposed a security knowledge operationalization approach for AI-assisted specification-driven development. The central argument is that security behavior should not remain implicit when functional behavior is explicitly specified for AI agents. Security knowledge must be represented as a structured, traceable, and verifiable artifact.

The Multilayer Specification Security Model connects system entities, threats, risks, requirements, implementation rules, controls, verification scenarios, and evidence. The Security Knowledge Transition Method describes how business and technical specifications can be transformed into a security-enriched generation contract.

Two empirical studies provide preliminary evidence. The hidden-oracle study showed that an LLM-based pipeline can recover broad security concerns from structured system context and produce a traceable candidate model. It achieved 100% relaxed coverage of oracle threats and risks, with weighted coverage of 72.73% for threats and 83.33% for risks. The backend generation study showed that security conditioning reduces hidden failures: modal failures decreased from 50 with no explicit security requirements to 42 with ASVS and 36 with the Multilayer Security Model. This corresponds to pass-rate improvements from 77.38% to 81.00% and 83.71%, and relative failure reductions of 16.00% and 28.00%.

The largest gains appeared in application-specific categories such as Business Logic and Admin Safety. This supports the claim that application-specific security modeling reduces security knowledge loss in AI-assisted specification-driven generation. Generic standards such as ASVS are useful, but they should be treated as security knowledge bases rather than final generation artifacts. The final artifact should bind requirements to concrete actors, resources, ownership rules, threats, controls, implementation rules, and verification expectations.